\documentclass{article}

\usepackage{arxiv}

\usepackage[utf8]{inputenc} 
\usepackage[T1]{fontenc}    
\usepackage{hyperref}       
\usepackage{url}            
\usepackage{booktabs}       
\usepackage{amsfonts}       
\usepackage{nicefrac}       
\usepackage{microtype}      
\usepackage{lipsum}
\usepackage{graphicx}
\usepackage{makecell}
\graphicspath{ {./images/} }

\title{Limitations of Online Play Content for Parents of Infants and Toddlers}

\author{Keunwoo Park, Subin Ahn, Mina Jung, You Jung Cho, Seulah Jeong, and Cheong-Ah Huh \\
Ordinary Magic Corp.\\
Seoul, Republic of Korea\\
\{keunwoo, subin, mina, oplm0318, seulah, cahuh\}@ordinarymagic.kr
}

\begin{document}
\maketitle
\begin{abstract}
Play is a fundamental aspect of developmental growth, yet many parents encounter significant challenges in fulfilling their caregiving roles in this area. As online content increasingly serves as the primary source of parental guidance, this study investigates the difficulties parents face related to play and evaluates the limitations of current online content. We identified ten findings through in-depth interviews with nine parents who reported struggles in engaging with their children during play. Based on these findings, we discuss the major limitations of online play content and suggest how they can be improved. These recommendations include minimizing parental anxiety, accommodating diverse play scenarios, providing credible and personalized information, encouraging creativity, and delivering the same content in multiple formats.
\end{abstract}


\section{Introduction} \label{sec:introduction}
Play is widely recognized as a crucial component for facilitating child development. Ginsburg~\cite{Ginsburg2007} underscores the significance of play in fostering brain development, bolstering self-assurance, and cultivating creativity and leadership skills. Furthermore, inadequate provision of a safe play environment and lack of nurturing relationships can potentially subject children to toxic stress, thereby compromising both the structural and functional development of their brains~\cite{Yogman2018}.

Parents may encounter challenges when engaging in play activities with their children regardless of their significance. Marshall~\cite{Marshall2018} notes that some parents express concerns about their perceived lack of creativity in devising play activities, often seeking assistance. Additionally, Gao et al.~\cite{Gao2021} analyzed prominent parenting topics on Reddit, revealing that \emph{play} emerged as one of the top 30 discussed topics. Their result suggests that discussions about parent-child play are prevalent in parenting discourse.

On the other hand, the internet plays an important role in solving other parenting problems. New parents get advice and information from search engines and use digital technologies to log their parenting life~\cite{Gibson2013}. Gao et al.'s~\cite{Gao2021} study showed that various parenting information is shared through one of the largest online communities, Reddit. Parents also create new social relationships through digital technologies~\cite{Toombs2018}.

In this context, the present study investigates parents' challenges when acquiring and utilizing play information online. Furthermore, the study discusses the possible solutions to address these obstacles. This research is limited to common online media, such as websites, blogs, and Social Network Services (SNS) like Instagram and YouTube. The term \emph{play information} or \emph{play content} denotes the contents that explain how to do a certain play or general information about baby play. For example, the content can be a video explaining creative plays using a rattle or a blog post about what kind of play is good for a 6-month-old child's gross motor development.

\section{Related Work}

\subsection{Importance of Child Play}
Numerous studies have consistently highlighted the importance of play in a child’s physical, cognitive, language, social, and emotional development. In particular, researchers focusing on early childhood play and development have emphasized the role of parents. Dhas et al.~\cite{Dhas2022} have noted that parents act as \emph{gatekeepers} in determining their children's play. In other words, parents’ perception and knowledge of play can significantly impact their children's play opportunities. Also, Tamis-LeMonda et al.~\cite{Tamis-Lemona2004} have shown that parental engagement in play with their children is related to their language and cognitive development. Additionally, Symons et al.~\cite{Symons2006} revealed that interactions during mother-child free play significantly relate to children’s theory of mind, a core element of social cognition and emotional development.

Despite the many studies highlighting the role of parents in play, especially during children's early years, there is a lack of empirical evidence regarding the difficulties they encounter in play and understanding of how parents address these challenges. This paper aims to identify parents' challenges while playing with their children and understand how they address them using the Internet.

\subsection{How the Internet Helps Parenting}

Many researchers have studied how the internet helps parenting regarding sources of information and communication channels. However, the Internet's role in parent-child play has been relatively less spotlighted in the CHI community. Marshall~\cite{Marshall2018} designed Digital Toybox, an online community where parents can share play ideas, which leads to helping other parents having difficulties playing with their children. To design such a tool, they interviewed and observed parents in a community program for local parents called `Little Monkeys.' From the study, they derived three requirements for the Toybox. First, it supported parents sharing the burdens of parenting. Second, it provides play activities that require a small cost. Third, it connected parents with kids of similar ages to enable the sharing of ideas. Although Digital Toybox is an early example of how an online community can support parent-child play, the author designed it based on the findings of offline behaviors, where the usage of popular online platforms in terms of play is relatively less studied.

Some studies focused on the relationship between popular online social media and parenting. Toombs et al.~\cite{Toombs2018} studied how sociotechnical systems help new parents. They found that new parents make new connections and get information about parenting and events for children. Also, online social support networks can help parents of postpartum depression in informational and emotional ways~\cite{Stana2019}. Gao et al.~\cite{Gao2021} showed that many parenting topics are actively discussed on one of the world’s largest online forums, Reddit. Morris~\cite{Morris2014} studied how mothers of young children use social network services and what information they seek. The studies above focused on general parenting. However, we think more research projects concentrating on play are necessary because play has a unique context compared to other domains of parenting.

\subsection{Technologies Supporting Parent-Child Relationship}

Many papers about technology support relationships between parents and children~\cite{Weimin2023, Stuckelman2023, Johnson2024, Musick2021}. Shin et al.~\cite{Shin2021} provided an overview of CSCW research projects supporting parent-child relationships. The difference between the stream of projects and our work is that we focus on the problems of online play content. This paper does not focus on parent-child interactions but on the relationship between parents and online play information.

It is worth mentioning that Wilkinson et al.~\cite{Wilkinson2018}'s work is close to ours. They studied a software application that facilitates parent-child real-world play. The application provided 150 play contents, and parents would see the contents and play with the children as the contents say. While Wilkinson et al. focused on one app, this paper considers common online media, such as Instagram and YouTube. Moreover, we target parents of children before two years old.

\section{Understanding How Parents Acquire Information About Play}

\begin{figure}
    \centering
    \includegraphics[width=0.75\linewidth]{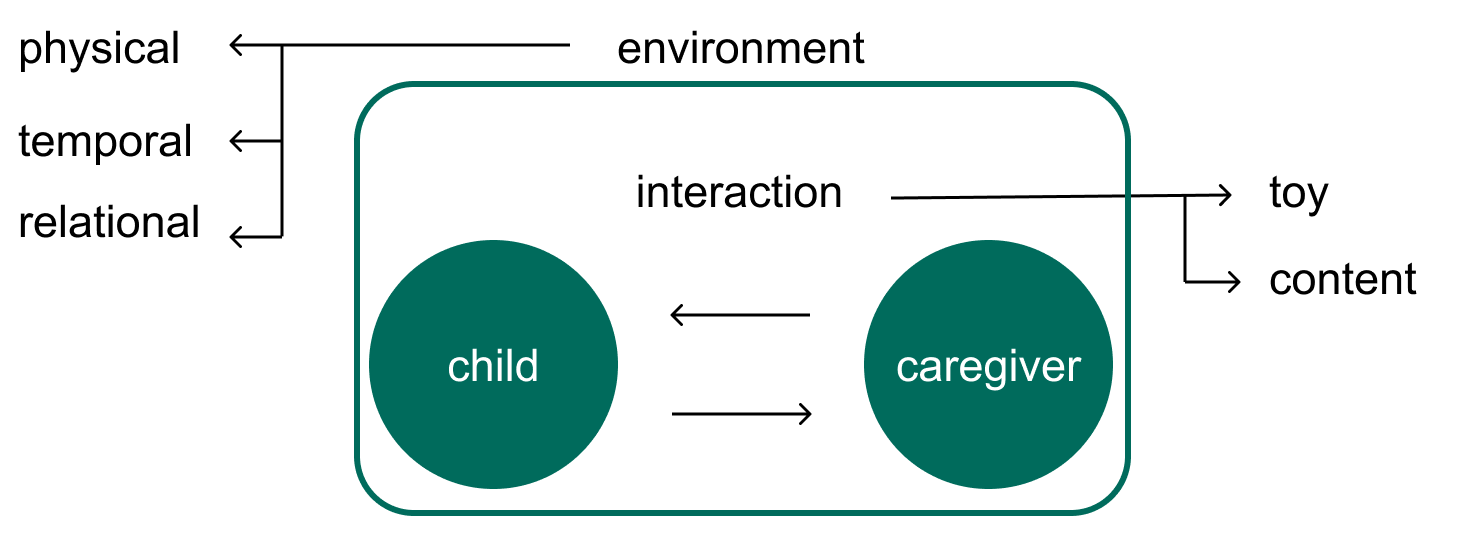}
    \caption{Our definition of the structure of play.}
    \label{fig:structure-of-play}
\end{figure}

We first aimed to understand how parents play with their children and how they utilize online content to overcome the challenges of playing with their children. We have outlined the concept of play, as illustrated in Figure~\ref{fig:structure-of-play}. When a caregiver and a child play together, the play should have a theme. The theme is limitless. It can be role-playing, tummy time, or any play. The theme defines how the caregiver and child interact. Mediums, such as toys, can be included to support the play. The surrounding environment can be divided into when and where they are playing together and what is the relationship between the caregiver and the child. The caregiver can be a parent, a grandparent, or a babysitter. We conducted online interviews for the understanding.

\subsection{Recruitment}

We wanted the children's ages of the participants to be as diverse as possible. Also, as mentioned earlier, we tried to recruit parents who experienced some difficulties playing with their children. Therefore, we sent a short survey asking the children's ages and how much they felt happiness and difficulty playing with their children to the parents with children less than 2 years old. When asking about happiness and difficulties, we used a five-point Likert scale. Then, we picked parents of diverse children of different ages who had difficulties playing. As a result, we recruited nine participants. Table~\ref{tab:presurvey} shows the information about the participants and their pre-survey answers. We compensated the participants with toys worth around 104 US dollars.

\begin{table}[]

\caption{The results of the pre-survey. Each column describes a) the age of each participant, b) the gender of each child, c) the age of each child in months, d) whether each child goes to a daycare center, e) how much each participant enjoys playing with their children, f) how much each participant faces difficulties in playing and g) the type of occupation of each participant. F, N, and P denote full-time, not working, and part-time, respectively. The answers to 'e' and 'f' were on a 5-point Likert scale.}

\label{tab:presurvey}
\centering
\begin{tabular}{llllllll}
\hline
   & a) age & \makecell{b) child \\ gender} & \makecell{b) child \\ age (months)} & d) daycare & \makecell{e) play \\ enjoyment} & \makecell{f) play \\ difficulty} & g) occupation \\ \hline \hline
p1 & 35 & M  & 7  & O  & 3  & 5  & F  \\ \hline
p2 & 39 & M  & 5  & X  & 5  & 3  & N  \\ \hline
p3 & 38 & M  & 4  & X  & 4  & 5  & N  \\ \hline
p4 & 34 & F  & 16 & X  & 4  & 2  & P  \\ \hline
p5 & 40 & F  & 2  & X  & 4  & 4  & N  \\ \hline
p6 & 34 & M  & 19 & O  & 4  & 4  & F  \\ \hline
p7 & 32 & F  & 6  & X  & 3  & 5  & N  \\ \hline
p8 & 31 & M  & 11 & O  & 5  & 4  & F  \\ \hline
p9 & 33 & F  & 13 & X  & 5  & 3  & F  \\ \hline
\end{tabular}
\end{table}

\subsection{Procedure}

We interviewed each participant online. Each interview took around 30 minutes. Detailed questions are listed in Table~\ref{tab:interview-questions}. The interviews were semi-structured so that interviewers could ask additional questions.

\begin{table}[ht]
\caption{The major questions for the interview.}
\label{tab:interview-questions}
\begin{tabular}{p{0.9\columnwidth}}
\textbf{Life} \\ \hline \hline
Please depict your ordinary weekdays and weekends. \\ \hline
What kinds of play do your children do most often? \\ \hline 
\\ 
\textbf{Play Style} \\ \hline \hline
What do you do when your children play? Do you let them play freely and observe? Or do you proactively lead them? \\ \hline
Do you plan the play? Or do you play spontaneously? \\ \hline
What do you do when the playtime does not go as you planned? \\ \hline
\\
\textbf{Toys and Environments} \\ \hline \hline
What kinds of toys do you prefer and why? \\ \hline
Where do you play most often? \\ \hline
\\
\textbf{Difficulty} \\ \hline \hline
What are the difficulties when playing with your children? How do you overcome these difficulties? \\ \hline
\\
\textbf{Content} \\ \hline \hline
Do you search for content to figure out what to play with the children? 

If so,

- Where and how often do you find them?

- What are your favorite sources and why?

- What are the characteristics of the contents that you wanted to try?

- Do you try to play with the children right after watching the play content or later?

- If there were contents that you wanted to try but did not, what was the reason?

If not,

- Why do you not search for such content?

- Are there any difficulties in playing with your children without such content?
\\ \hline 
\\
\textbf{Development} \\ \hline \hline
Were there worrying moments when you questioned whether your children were growing well? \\ \hline
What are the moments when you felt your children’s growth? \\ \hline
Do you record your children’s growth? If so, how do you record and why? If not, do you want to record it? \\ \hline

\end{tabular}    
\end{table}

\subsection{Analysis Method}

We derived findings from the interviews throughout two affinity diagram sessions. Two researchers transcribed the whole interview. Then, all the researchers read the transcripts or listened to the recordings of the interviews. Each researcher itemized interesting quotes from the interviews for affinity diagrams. For the first round of sessions, we organized items per each participant. Then, all the researchers grouped the items on their own. Next, we looked through all the groups together and adjusted the items and the groups. We found 136 groups in this round, considered as \emph{observations}. Each observation was converted as an item for the next affinity diagram round. All the researchers also participated in the second round and grouped the items. When we filtered out only the groups related to play content, ten of them remained, which are our \emph{findings}. Everyone agreed on the result of each round.

\section{Findings}

The findings can be divided into two categories. The first concerns the challenges in parent-child play in general. The second concern is the relationship between parents and online play content. Understanding both categories is essential for designing effective online content that enhances parent-child play.

\subsection{Challenges of Play}

The participants had difficulty with their children not following the play as they had planned. They were apprehensive and stressed when they could not observe expected behaviors or reactions. $P4$ told us about difficulties coming from the gap between her expectation and the child's response by saying, ``There are often times that my child did not follow my guide when I want to teach her something.'' This phenomenon made some participants think play guide content is useless. $P6$ said, ``I wanted to follow a play guide, but my child was not interested in the play. In such cases, I feel like the guides restrict me. It is too hard for me to utilize play guides.'' $P1$ said, ``I read some play guides in a mobile app and tried to play as they said. However, my child did not follow my lead. So, I saw the mobile app less and less.''

Similarly, the participants struggled to find their children's interests. Also, they were embarrassed when their children did not show interest in the toys that they thought their children would. $P5$ said, ``It is hard to determine whether my child likes or dislikes a toy based on the reaction.'' and $P3$ said, ``My child liked a toy with a mirror for a while, but he has no more interest in it.'' In addition, we found out that is hard to induce interest from children. $P4$ said, ``I struggled when my child showed less interest than I expected because I didn't know how to spark their enthusiasm.''

The lack of reaction from children makes the play unilateral, which frustrates some participants and causes them to lose motivation to play. This might happen more often when a child is in a period when they cannot show significant reactions during interaction with their parents. $P1$ said, ``It was difficult when I played with my child unilaterally. I was curious whether my child was recognizing my effort.''

The participants were pressured to perform various plays. This issue has two aspects. First, the time spent with children is too long compared to the number of possible plays. $P7$ said, ``The child is awake longer as she grows. So, I struggle to find a play for her. My plan is over, and she is no longer interested in the plays that I present.'' Second, it is hard to come up with creative play. $P1$ said, ``Honestly, it is too hard. There is no textbook for play. I have to be creative to play with my children.'' Also, $P8$ said, ``I have the pressure to achieve both quantity and quality of play.'' While parents want to do various creative plays, $P9$ mentioned that it is hard to find play content. This finding is related to Marshall~\cite{Marshall2018}'s work, as they also pointed out the parents were worried about their lack of play-related creativity.

\subsection{Online Play Content} \label{sec:difficulties}

All the participants acquired play and parenting content from the internet. However, the participants thought SNS content was unreliable and had their strategies for filtering reliable content. $P8$ followed accounts of child health and development experts, and $P6$ did not credit accounts that sold products. $P1$ cross-checked content from non-experts. She said, ``There are common plays for children in a certain period that multiple accounts post. Then, I think the plays are necessary for my kids.'' However, some participants struggled with too many resources on the internet. $P5$ said, ``It is hard to find content that fits my kids. There are too many things, and it feels like I have to do everything.''

Some participants were stressed when they saw other kids on SNS because they sometimes compared their children with those on SNS. Especially, $P1$ said, ``It seems like the development of kids on the internet is faster than average. Watching them makes me have more expectations for my child. I know that some kids grow slower and eventually grow up well. However, the internet pressures me to do more.''

Although participants get a lot of play content, it is hard to utilize all the content because they do not have time to practice and get used to new plays. $P3$ said, ``I realized that I could not utilize 100\% of all the content that I saw because I cannot see the content at the same time I play with my child.'' $P3$ also said, ``It is hard to react to my child's actions when I do new play because I'm not used to it. Therefore, although I learn new plays, I repeat the play that I already know.'' As a result, participants preferred easy-to-follow plays, not effective plays. $P4$ said, ``Among the plays I saw on the internet, I try plays that are easy to follow and what my child would like.''

Some participants worried about the safety of internet content. $P3$ wanted to do a play she saw on SNS, but she said, ``I was worried about the safety. I was unsure whether my child was developed enough for the play.''

Similarly, the traditional media was important in terms of credibility. Some participants got information from acquaintances who have children of similar age to theirs. $P7$ said, ``I asked other parents who I met in the community center.'' and $P9$ said, ``I asked many questions to a parent I met in co-parenting who has three kids and is a teacher.'' Books and academic papers were not uncommon. $P8$ said, ``I prefer to get child development information from papers and books.'' $P2$ said, ``I read parenting books when my kid is asleep, and I could find the right parenting direction.'' However, the problem with the books was that they provided too general information. $P6$ said, ``I'm an education major. I often read textbooks from my undergrad classes, but I found that theory and reality are vastly different.'' $P8$ said ``I searched books for specific information in a specific context. However, I could only find general and abstract information because the books covered various ages.''

The participants utilized video and static (i.e., text and image) content in different contexts. Some participants preferred video content because it was easier to catch small details of what parents should do during plays. $P3$ said, ``I prefer reading text for other content, but I prefer video play content because it is about body movement.'' Also, $P1$ preferred video because she can glance at it, which does not require full attention. Some participants, such as $P3$ and $P8$, preferred static content because it was faster to get the necessary information. It is related to the parenting environment. Parents might not want to expose their children to smartphones; in such cases, they have to find the necessary information and leave the phone quickly. $P8$ said, ``I minimize using my smartphone before and during playtime because my kid wants to see the smartphone when I use it.''

\section{Discussion: The Limitations of Online Play Content and Possible Solutions}

We believe that the challenges associated with play are largely due to the limitations of online content. Our participants struggled to find solutions to various problems, although they utilized online content. Consequently, this section examines the shortcomings of online play content in light of the broader challenges related to play.

Our data show that online content makes parents anxious when showing well-playing babies. The participants were exposed to content showing babies being developed faster than average, pushing them to have higher expectations for their children. It eventually stressed the participants because children cannot meet their parents' expectations perfectly. Therefore, we think when play content exposes playing scenes of other children to parents, it should use children in similar developmental stages to the user's children because parents get anxious when they think their children are developing slower than others. Also, the content should explain that children might not follow all the plays that parents suggest. At the same time, it should guide parents on how they can help their children to enjoy certain play.

Based on our study, we argue online content might not be helpful for parents responding to various scenarios during play. We found out the parents struggled when their children did not respond as expected. Also, they struggled to find children's interests and were frustrated when children did not show enough reaction. We think online content could handle these situations by providing how parents can respond to various scenarios while doing certain play. For example, play content would explain how to do a play. Better content could guide parents in various scenarios, such as when children suddenly show interest in another toy. In such cases, parents might be confused about whether to let the child move on to another toy or to induce the child to focus on the current toy.

Play is a creative activity, but online content is not supporting parent's creativity, according to our interviews. The participants were pressured to prepare various plays and were frustrated by repetitive plays. We think this might be related to the finding that parents cannot fully utilize their acquired content. Although there is a lot of online play content, they might act as entertaining content, not valuable guides. Therefore, we think play content should describe a particular play and allow the parents to be creative. For example, the content might be designed to help memorize the play so they can recall it during playtime. Also, content can explain how parents can vary a play.

Our results show that online content is not credible and that it is not tailored information. We found that parents become more skeptical about child-related information on the internet. Our participants trusted information from experts and acquaintances with children but did not trust other parents on SNS, especially commercialized accounts. However, credibility is not the only factor. The participants also struggled to find the information they needed from highly trusted media, such as books and papers. Books and papers provide general knowledge, but parents want information for a specific situation. The solution would be an application that provides tailored information based on the child's development process from credible sources.

It could be helpful if the same content should be provided in dynamic (e.g., video) and static (e.g., text and images) forms because they are helpful in different contexts. Since play is a physical activity, parents might want to get detailed guides on how they should act and talk during play. However, there are times when parents should use smartphones for a short time, such as when they are with their children. Static contents are helpful in such cases because parents can quickly read content and get back to their children. We think the current online media is unsuitable for such needs because they provide either video or text. A hybrid media would be appropriate. 

\section{Limitations and Future Work}

This research only involved South Korean women due to the difficulty of recruiting male major caregivers in our recruiting pool. Responses might differ based on culture and gender. Therefore, we plan to include more diverse cultures and genders in future work. Moreover, we will design and implement an application that resolves the limitations of online content and evaluate its effectiveness.

We suggest such an application should be interactive because the guides should consider various play situations, and linear content -- like recipes -- is inappropriate. We think the application should first provide base play content and take input from the parents about how the play went. Then, the application could provide appropriate consecutive content based on the user input. However, it also should instruct the parents to use the application before or after the play. According to Wilkinson et al.~\cite{Wilkinson2018}, parents using applications while they are with children might distract the children's attention.

\section{Conclusion}

This work finds challenges for parent-child play and discusses the limitations of online play content. We interviewed nine parents of diverse children of different ages and derived ten findings. Based on the findings, we discussed the limitations of the current online content and possible solutions. We believe that play content should not make parents anxious, help them respond to various play situations, and let them be more creative. This research would be the groundwork for designing software to help parents play well with their children.

\bibliographystyle{unsrt}  
\bibliography{references}  

\end{document}